\documentclass[jgrga]{agutex}

  \usepackage[dvips]{graphicx}
\authorrunninghead{OKA ET AL.}

\titlerunninghead{ISLAND SURFING DURING RECONNECTION}

\authoraddr{M. Oka, T. D. Phan,
Space Sciences Laboratory, University of California Berkeley, 7 Gauss Way Berkeley, CA 94720-7450, USA. (moka@ssl.berkeley.edu)}

\authoraddr{M. Fujimoto, I. Shinohara
Institute of Space and Astronautical Science, Japan Aerospace Exploration Agency, 3-1-1 Yoshinodai, Sagamihara, Kanagawa, 229-8510, Japan.}

\begin{document}

%% ------------------------------------------------------------------------ %%
%
%  TITLE
%
%% ------------------------------------------------------------------------ %%

\title{`Island Surfing' Mechanism of Electron Acceleration\\
During Magnetic Reconnection}

% e.g., \title{Terrestrial ring current:
% Origin, formation, and decay $\alpha\beta\Gamma\Delta$}
% You may use \\ to break the title over several lines.

%% ------------------------------------------------------------------------ %%
%
%  AUTHORS AND AFFILIATIONS
%
%% ------------------------------------------------------------------------ %%

\authors{M. Oka, \altaffilmark{1}
M. Fujimoto, \altaffilmark{2}
I. Shinohara, \altaffilmark{2}
T. D. Phan \altaffilmark{1}
}

\altaffiltext{1}{Space Sciences Laboratory, University of California Berkeley, Berkeley, California, USA.}
\altaffiltext{2}{Institute of Space and Astronautical Science, Japan Aerospace Exploration Agency, Sagamihara, Kanagawa, Japan.}

%Use \author{\altaffilmark{}} and \altaffiltext{}

% \altaffilmark will produce footnote;
% matching altaffiltext will appear at bottom of page.
% May use \\ to start a new line.

%\authors{R. C. Bales, \altaffilmark{1}
%E. Mosley-Thompson, \altaffilmark{2} R. Williams, \altaffilmark{3}
%and J. R. McConnell\altaffilmark{4}}

%\altaffiltext{1}{Department of Hydrology and Water Resources,
%University of Arizona, Tucson, Arizona, USA.}

%\altaffiltext{2}{Department of Geography, Ohio State University,
%Columbus, Ohio, USA.}

%\altaffiltext{3}{Department of Space Sciences, University of
%Michigan, Ann Arbor, Michigan, USA.}

%\altaffiltext{4}{Division of Hydrologic Sciences, Desert Research
%Institute, Reno, Nevada, USA.}

%% ------------------------------------------------------------------------ %%
%
%  ABSTRACT
%
%% ------------------------------------------------------------------------ %%

% >> Do NOT include any \begin...\end commands within
% >> the body of the abstract.

\begin{abstract}
One of the key unresolved problems in the study of space plasmas is to explain the production of energetic electrons as magnetic field lines `reconnect' and release energy in a exposive manner. Recent observations suggest possible roles played by small scale magnetic islands in the reconnection region, but their precise roles and the exact mechanism of electron energization have remained unclear. Here we show that secondary islands generated in the reconnection region are indeed efficient electron accelerators. We found that, when electrons are trapped inside the islands, they are energized continuously by the reconnection electric field prevalent in the reconnection diffusion region. The size and the propagation speed of the secondary islands are similar to those of islands observed in the magnetotail containing energertic electrons.

\end{abstract}

%% ------------------------------------------------------------------------ %%
%
%  BEGIN ARTICLE
%
%% ------------------------------------------------------------------------ %%

% The body of the article must start with a \begin{article} command
%
% \end{article} must follow the references section, before the figures
%  and tables.

\begin{article}

%% ------------------------------------------------------------------------ %%
%
%  TEXT
%
%% ------------------------------------------------------------------------ %%

\section{Introduction}

The origin of energetic electrons of up to MeV in the Earth's magnetotail has been an outstanding problem for decades \citep[e.g.][]{sarris76, terasawa76}. Although many theories consider magnetic reconnection as the mechanism of explosive energy release phenomena in the Earth's magnetosphere,  the exact mechanism by which high energy electrons are produced is still unclear. 

A statistical survey by the Geotail spacecraft shows that accelerated and/or heated electrons are more likely to be detected around the outflow regions somewhat away from the center of magnetic reconnection \citep{imada05}. It was suggested that electrons pre-energized at the X-lines further gain energy in the flux pile-up region especially in the earthward side of the X-line.  While a Cluster event at the near-Earth reconnection site also confirms this scenario \citep{imada07}, another event by the Wind spacecraft in the distant magnetotail observed increasing flux of energetic electrons with decreasing distance from the X-line, with no clear evidence for a flux pile-up region in the outflow jet, suggesting that the region around the X-line is the dominant source of energetic electrons \citep{oieroset02}. The discrepancy between the two events is not well understood so far.

On the other hand, recent measurements by the Cluster spacecraft pointed out that magnetic islands may be an important agent responsible for the production of energetic electrons \citep{chen08, retino08}. The main features of the observed islands are that they are small - of the order of ion inertia length - and that they move fast, close to the Alfv\`{e}n speed. These features suggest that they are the magnetic islands that are naturally formed as a consequence of secondary tearing instability of a thin current sheet \citep[][and references therein]{eastwood07}.

As for theoretical models of electron acceleration within magnetic islands, a contracting motion of volume filling magnetic islands has been considered \citep{drake06}. Because of the contracting motion, inductive electric fields are created at each end of the islands and particles that are fast enough to circulate inside the islands can receive repetitive `kicks' from the electric fields. 
%This time-dependent model is analogous to the energy increase of a ball reflecting between two converging walls, namely the first order `Fermi' process. 
However, the presence of volume filling islands and their contracting motion are difficult to verify observationally and have not been reported.

In this Paper, we present an alternative scenario of electron acceleration by performing 2D particle-in-cell (PIC) simulations of magnetic reconnection.  We found that electrons are trapped in a secondary magnetic island so that they can be continuously energized by the reconnection electric field in the diffusion region.

\section{Simulation}

%The simulation configuration is the same as in our recent study of multi-island coalescence and associated acceleration of electrons \citep[][]{oka10b}. 
We utilize a two and half dimensional, fully relativistic PIC code \citep{hoshino87, shinohara01}. The initial condition consists of two Harris current sheets. The anti-parallel magnetic field is given by $B_y/B_0=\tanh((x-x_R)/D)-\tanh((x-x_L)/D)-1$ where B$_{\rm 0}$ is the magnetic field at the inflow boundary, $D$ is the half-thickness of the current sheet and $L_x$ and $L_y$ are the domain size in $\mathbf{\hat{x}}$ and $\mathbf{\hat{y}}$ direction, respectively. $x_L$=$L_x/4$ and $x_R$=3$L_x/4$ are the $x$-positions of the left and right current sheet, respectively.  Periodic boundaries are used in both directions. The ion inertial length $d_i$ is resolved by 25 cells. D=0.5$d_i$ and $L_x=L_y$=102.4$d_i$. The inflow, background plasma has the uniform density of $N_{B0}$=0.2$N_0$ where $N_0$ is the density at the current sheet center. The ion to electron temperature ratio is set to be $T_i/T_e$=5 for the current sheet and $T_i/T_e$=1 for the background. The background to current sheet temperature ratio $T_{bk}/T_{cs}$=0.1. The frequency ratio $\omega_{\rm pe}/\Omega_{\rm ce}$=3 where $\omega_{\rm pe}$ and $\Omega_{\rm ce}$ are the electron plasma frequency and the electron cyclotron frequency, respectively. The ion to electron mass ratio $m_i/m_e$=25 and the light speed $c$ is 15V$_A$ where V$_A$ is the Alfv\'{e}n speed defined as B$_0/\sqrt{4\pi N_0 m_{\rm i}}$. We used average of 64 particles in each cell. 297 particles per cell represents the unit density. 

No magnetic field perturbation is imposed at the beginning so that the system evolves from a tearing mode instability due to particle noise. Note that conventional simulations initiate magnetic reconnection by a small magnetic field perturbation in order to save computational time \citep[e.g.][]{birn01}. Our simulation setup does not use such trigger so that the system does not produce any large initial velocity flows resulting from a lack of pressure balance. 

\begin{figure}[t]
\includegraphics[width=80mm]{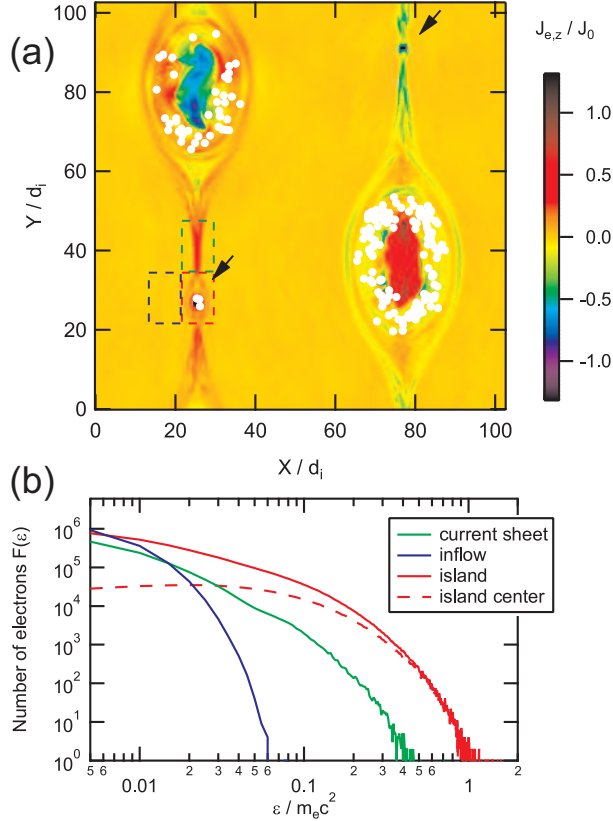}
\caption{(a) The out-of-plane component of the electron current density at $\Omega_{\rm ci}$t=115. Superposed on the image are the locations of the most energetic electrons ($\varepsilon>$1.4m$_ec^2$). (b) Electron energy spectra integrated over the rectangular regions indicated in (a) (solid curves). The dashed curve shows the spectrum at a square region centered at  ($x,y$)=(25.6d$_{\rm i}$, 26.4d$_{\rm i}$) with each side 1.6d$_{\rm i}$. \label{fig:csze}}
%(red curve), ($x,y$)=(25.6d$_{\rm i}$, 36.8d$_{\rm i}$) (green curve), and ($x,y$)=(15.2d$_{\rm i}$, 26.4d$_{\rm i}$) (blue curve). Each square side is 1.6d$_{\rm i}$. \label{fig:csze}}
\end{figure}

Figure \ref{fig:csze}a shows a snapshot of the out-of-plane component of electron current density $J_{e,z}$ obtained when the system is well developed ($\Omega_{\rm ci}$t=115). The diffusion regions are represented by the localized current densities in a thin current sheet, and embedded within the diffusion regions are the small scale secondary islands (highlighted by the arrows).  Superposed on the plot are the locations of the most energetic electrons ($\varepsilon>$1.4m$_ec^2$, indicated by the white circles). Most of the energetic electrons are located within the large islands but some electrons do exist in the small, secondary island of the left current sheet. 

%Note that the large islands are due to coalescence of initial magnetic islands generated by the initial tearing instability and are not responsible for the energetic electrons in the small islands. While electron energization processes in the large islands are fully described in our separate paper, this paper focuses on the energization of the electrons in the small islands.

Those particles within the large islands are energized through the so-called multi-island coalescence resulting from the non-linear evolution of the tearing mode instability. A variety of different energization mechanisms has been found and they are fully described in our companion paper \citep{oka10}. On the other hand, the energization of particles within the small islands are not the consequence of multi-island coalescence and the physics involved are entirely different, as will be shown below. Hereafter, we focus on the electron energization process within the small scale, secondary islands.

Figure \ref{fig:csze}b shows the electron energy spectra obtained at the time of Figure \ref{fig:csze}a. It is evident that, at $\Omega_{\rm ci}$t=115, the highest energy electrons in the diffusion region are all confined in the small magnetic island, particularly at the very center of the island (red curves). This island overlaps the electron-scale diffusion region (or the so-called `inner diffusion region') where electron acceleration should be efficient. For a reference, we also plotted a spectrum for the immediate upstream (blue curve) and the downstream of this region (green curve).

%Attached to this region are the elongated electron Alfv\`{e}nic flow exist. Such region does not produce much non-thermal electrons \citep{oka10} (green curve). For a reference, we also plotted a spectrum for the inflow region (blue curve).
% However, this doe not mean the island is the only agent that brings the electron energy up to $>$1.4m$_ec^2$ because the whole diffusion region is not time stationary. There is a possibility that the highest energy electrons might have been pre-energized at the diffusion region center when the island has not been appeared.

%sThe black curve shows the spectrum integrated over the rectangular region of the left current sheet in Figure \ref{fig:csze}a. It consists of a cold thermal component of the inflow region and a hot thermal component created within the diffusion region. While the diffusion region outside of the secondary island can energize electrons (green curve), the higher energy electrons are produced within the secondary island (red curve). 
%As theThe heating takes place in the In addition to the cold thermal distribution of the incoming plasma, there exists another hot distribution which tail extends as high as $>$1.0m$_ec^2$. The spectral form is similar to the results of previous studies \citep[e.g.][]{hoshino05, pritchett08}. 

\begin{figure*}
\includegraphics[width=170mm]{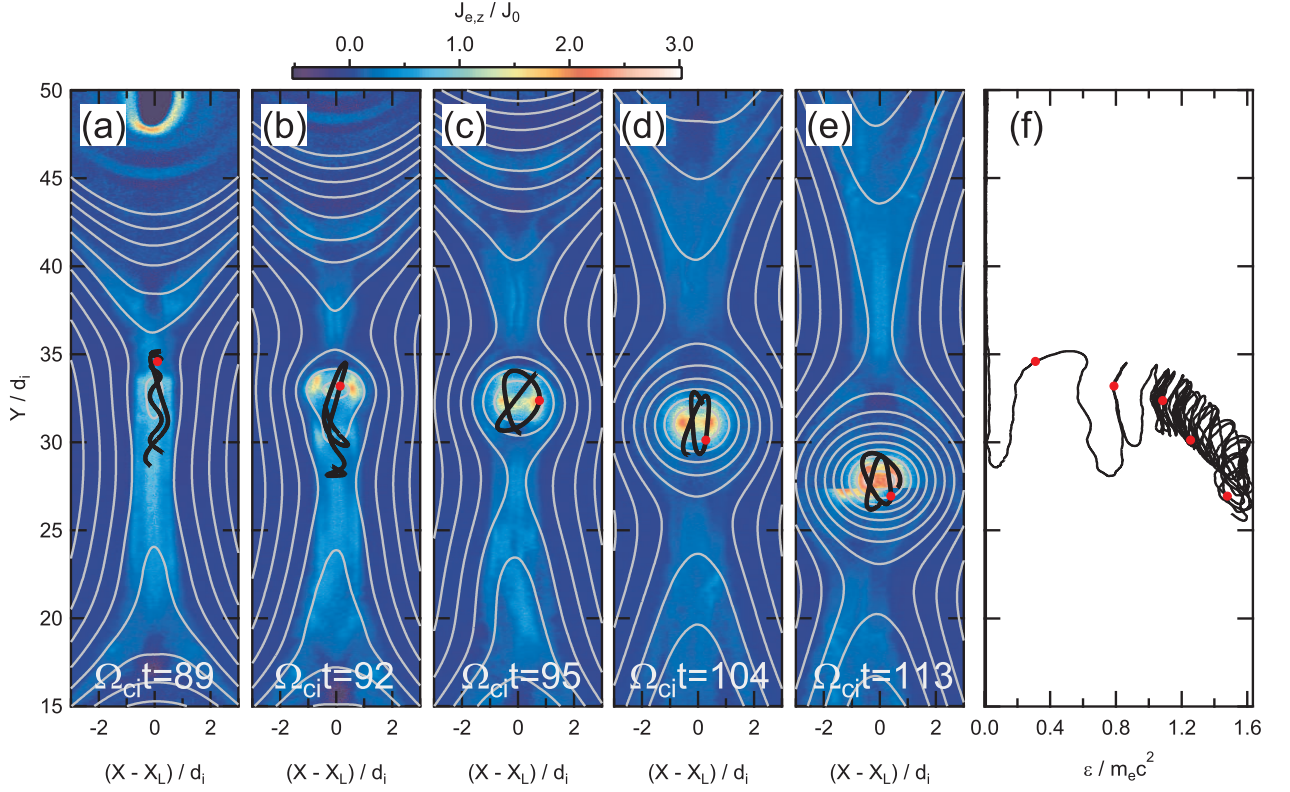}% Here is how to import EPS art
\caption{\label{fig:HL-overview} The trajectory of the most energetic electron during (a) 86.7$\leq\Omega_{\rm ci}t\leq$90.7, (b) 90.7$\leq\Omega_{\rm ci}t\leq$93.3, (c) 95.0$\leq\Omega_{\rm ci}t\leq$95.6, (d) 103.7$\leq\Omega_{\rm ci}t\leq$104.3, (e) 112.4$\leq\Omega_{\rm ci}t\leq$113.0. The background images are the electron current density at (a)  $\Omega_{\rm ci}$t=88.7 (b) $\Omega_{\rm ci}$t=92.0, (c) $\Omega_{\rm ci}$t=95.3, (c) $\Omega_{\rm ci}$t=104.0, and (d) $\Omega_{\rm ci}$t=112.7. In (e), the whole trajectory (0$\leq\Omega_{\rm ci}t\leq$115) is shown in a format of $y$ versus $\varepsilon$. Each red mark shows the time of the background image. }
\end{figure*}

In order to clarify the history of electron acceleration, we followed the trajectory of the highest energy electron in the rectangular box in Figure \ref{fig:csze}. In Figure \ref{fig:HL-overview}, we divided the trajectory into five segments and plotted over the image that shows the out-of-plane component of the electron current density. Note that the simulation itself is two-dimensional in $x$ and $y$ but the displacement $\Delta z$ can be calculated by integrating $v_z$ over time. The background images are the snapshots taken from each segment so care should be taken when comparing the trajectory with the images. The $y$ position as a function of energy $\varepsilon$ is also shown in the right hand panel.

The secondary magnetic island starts to appear at $\Omega_{\rm ci}$t$\sim$89 (Figure \ref{fig:HL-overview}a). 
% and the resultant perturbation in the diffusion region scatters the particle so that it stays within the diffusion region to receive sustantial amount of energy. 
The electron acceleration seen up until $\Omega_{\rm ci}$t$\sim$89 is what is typically seen in the inner diffusion region. Had the electron left the diffusion region upward soon after $\Omega_{\rm ci}$t$\sim$89, it would not have become one of the most energetic electrons in the system. In reality, it was deflected downward back to the diffusion region by the Lorentz force due to magnetic field of the emerging island. Then, a clear indication of the interaction between the electron and the secondary island is seen at $\Omega_{\rm ci}$t$\sim$92 (Figure \ref{fig:HL-overview}b). A distinct change of the electron behavior starts at $\Omega_{\rm ci}$t$\sim$95. The electron is trapped in a well developed island (Figure \ref{fig:HL-overview}c).  The island continues to develop so that the magnetic island becomes as large as $\sim$5d$_{\rm i}$ although electron current is rather localized of the order of 2d$_{\rm i}$ (Figure \ref{fig:HL-overview}d). During this later phase, the electron continues to gain energy constantly (Figure \ref{fig:HL-overview}f).

The behavior of the electron can also be studied from the time profiles of the electromagnetic field felt by the particle (Figure \ref{fig:HL-field}b). Before being trapped by the island, the electron is rapidly energized by the strong ($\sim$0.1V$_{\rm A}$B$_{\rm 0}/c$) reconnection electric field. We confirmed that the work done by the E$_z$ component, i.e.  (-e)$\int v_zE_z$ dt, dominates over the work done by the E$_x$ and E$_y$ components, i.e.  (-e)$\int v_xE_x$ dt and (-e)$\int v_yE_y$ dt. 

%In the diffusion region, a secondary island starts to appear and the particle orbit is modulated during the island appearance (95$\leq\Omega_{\rm ci}t\leq$100, Figure \ref{fig:HL-overview}b). During this period, the energy increases slightly and reaches $\varepsilon\sim$1.2m$_e$c$^2$. 
Now, our interest goes to the next period during which the energy increases almost constantly until $\Omega_{\rm ci}$t$\sim$110 (Figure \ref{fig:HL-field}a). The energy increment is about 0.3m$_ec^2$ and the displacement along the $z$ direction is quite large, $\Delta z\sim$ 110d$_{\rm i}$. What is important here is that the island is expanding rather than contracting during the energization (Figure \ref{fig:HL-overview}). Therefore, we must explore energization mechanism other than the contracting island mechanism \citep{drake06}. After about $\Omega_{\rm ci}$t$\sim$110 when the island reaches the edge of the diffusion region, the energy remains almost constant at $\varepsilon\sim$1.5m$_{\rm e}$c$^2$.% The island moves with the speed $\sim$0.3V$_{\rm A}$.

The source of energy can again be studied from the profile of the electric field felt by the particle (Figure \ref{fig:HL-field}b). The $x$ and $y$ components of the electric field (to be discussed in more detail below) are oscillating so that the net energy gain from these components should be small. Here, we focus on the $z$ component of the electric field, E$_z$, which is almost equivalent to the reconnection electric field. It remains, on average, constant at $E_z\sim$0.025V$_{\rm A}$B$_{\rm 0}/c$ until $\Omega_{\rm ci}$t$\sim$110 and then decreases to zero afterwards. Let us briefly check if this E$_z$ can explain the energy increment $\Delta\varepsilon \sim$0.3m$_{e}$c$^2$. Generally, the energy increase $\Delta \varepsilon$ is estimated as
\begin{equation}
\frac{\Delta \varepsilon}{m_{\rm e}c^2} = \left(\frac{\Omega_{\rm ce}}{\omega_{\rm pe}}\right)^2 \left(\frac{cE}{V_{\rm A}B_{\rm 0}}\right) \left(\frac{\Delta z}{d_{\rm i}} \right)
\end{equation}
where $\Omega_{\rm ce}/\omega_{\rm pe}$=1/3 in our simulation setup. For the particle we analyzed, it felt reconnection electric field E$_z$$\sim$0.025V$_{\rm A}$B$_{\rm 0}/c$ and was displaced $\Delta z\sim$110d$_{\rm i}$ during the period 100$\leq\Omega_{\rm ci}t\leq$110 so that the energy increment should be $\Delta \varepsilon\sim$0.3m$_{\rm e}c^2$ in agreement with the above measured value.  From these arguments, we conclude that the electron is energized by the reconnection electric field E$_z$ while being trapped within the secondary island. 

One may point out a possibility of adiabatic heating because the magnetic field magnitude shows gradual increase during the energization (Figure \ref{fig:HL-field}c). However, this is not the case. The spatial scale of the particle motion is comparable to the island size so that the trajectory is decoupled from the magnetic field lines, indicating that the electron is highly non-adiabatic (Figure \ref{fig:HL-overview}c-e). We also confirmed that the magnetic moment $\mu$ of this particle is not constant during the energization (not shown).

The main feature of the secondary magnetic island within a diffusion region is that it contains an electrostatic potential well. In the diffusion region, ions are unmagnetized while electrons are more magnetized so that
%secondary island structures are supported by the electron current in the out-of-plane direction, leading to the charge separation between ions and electrons. As a result, 
the in-plane, polarization electric field E$_{\rm p}$ is generated and converge toward the center of the magnetic island.

Figure \ref{fig:HL-pots}a and Figure \ref{fig:HL-pots}b respectively shows the potential and the electromagnetic field structure at and around the secondary magnetic island shown in Figure \ref{fig:HL-overview}c. 
% This is when the electron has been under constant energization. It is evident from the figure that
It is evident that the potential well exists in the magnetic island and that its depth is of the order of the kinetic energy of the reconnection outflow $\Phi\sim$m$_{\rm i}$V$_{\rm A}^2$. 
%The spatial scale of the potential, however, is only of the order of d$_i$ so that ion motion is not affected by the structure. Electrons, on the other hand, can interact with this structure because of their small gyro-radii. 
%It is worth emphasizing here that the accelerated electron energy is much larger than the potential energy of the secondary magnetic island. 
%Also, in retrospect, the oscillatory feature in Figure \ref{fig:HL-overview}f,g was due to electron motion within this potential structure.
The deep potential well is represented by the large, bipolar signature in the E$_{\rm y}$ profile. A small bipolar signature also appears in the otherwise flat profile of E$_{\rm z}$ because of the motion of the entire island with the speed $\sim$0.3V$_{\rm A}$ directed toward the negative $y$-direction. This structure led to the oscillatory profile of the E$_{\rm z}$ felt by the particle (Figure \ref{fig:HL-overview}f). Note again that the net energy gain is positive because of the E$_{\rm z}$ offset due to the reconnection electric field.

\section{Discussion}

\begin{figure}
\includegraphics[width=80mm]{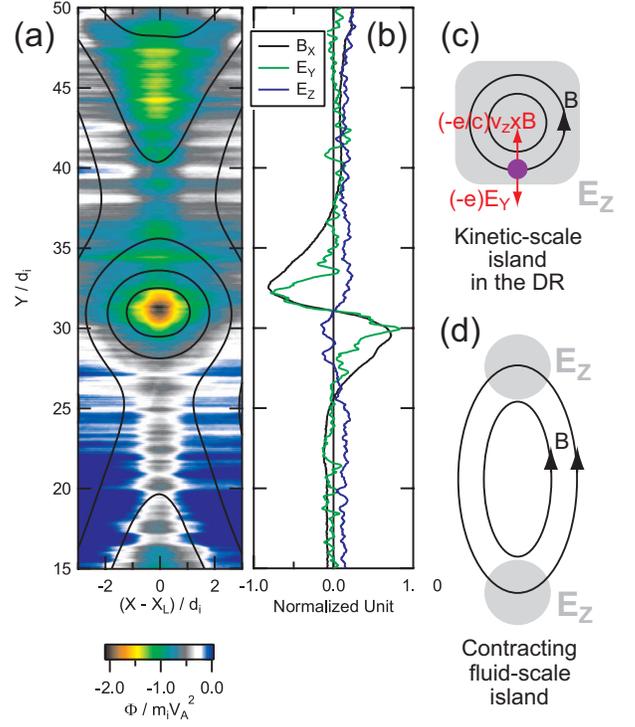}
\caption{(a) The electric potential at $\Omega_{\rm ci}$t=104. The contour shows the magnetic field lines. (b) A 1D cut through $x=x_L$ of B$_{\rm x}$/B$_{\rm 0}$, cE$_y$/V$_{\rm A}$B$_{\rm 0}$, and cE$_z$/V$_{\rm A}$B$_{\rm 0}$. (c) A schematic illustrations of a secondary island in the diffusion region (DR) and the force balance of an electron. (d) An illustration of a contracting island.\label{fig:HL-pots}}
\end{figure}

We found an energization mechanism by which electrons are trapped by a small scale magnetic island so that they are continuously energized by the reconnection electric field within the diffusion region. This is similar to the `surfing' mechanism by which electrons are accelerated by the reconnection electric field while `surfing' along the layer of the in-plane, polarization electric field \citep{hoshino05}. In order to discriminate the two mechanisms, we term the newly found mechanism as the `island surfing'. Let us now consider the `island surfing' in a similar way as discussed by \cite{hoshino05}. As an example, we consider an electron that is located along the current sheet center ($x$=$x_L$) but slightly away from the island center (Figure \ref{fig:HL-pots}c). From the symmetry of the electromagnetic fields at and around the magnetic islands, the following arguments can be applied to other electrons without losing generality. The $y$-component of the equation of motion reads %$dp_y/dt = (-e)E_y + (-e/c)(v_zB_x - v_xB_g)$, 
\begin{equation}
\frac{dp_y}{dt} = (-e)E_y + \left(-\frac{e}{c}\right)(v_zB_x - v_xB_g)
\end{equation}
where $p_y$=$m_ev_y$ is the electron momentum,  E$_y$ on the right-hand side can be replaced by the polarization electric field E$_p$ produced within the magnetic island,  and the guide field $B_g$=0 in the present case. If the Lorentz force $F_B$=(-e/c)$v_zB_x$ is larger than the electric force $F_E$=$(-e)E_p$, the electron is trapped and will gain energy from the reconnection electric field $E_z$. 
From the trapping condition, we get 
\begin{equation}
|v_{z}|>c\frac{E_p}{B_x} \sim \left(\frac{d_{\rm i}}{l}\right)V_A
\end{equation}
indicating that the `island surfing' mechanism requires a pre-acceleration to be kept trapped within an island. In general, the polarization electric field $E_p$ is expressed as%$E_p/(B_0V_A/c) = d_{\rm i}/l$, 
\begin{equation}
\frac{E_p}{B_0V_A/c} = \frac{d_{\rm i}}{l}
\end{equation}
where $l$ is the thickness of the  potential structure and may be estimated as an intermediate scale between ion inertia and the electron inertia length, $l=c/\sqrt{\omega_{\rm pe}\omega_{\rm pi}}$ \citep{hoshino05}. In the present case, $l$ is measured from the radius of the secondary island and is of the order of the ion inertial length. This is not unrealistic because the size (diameter) of the magnetic islands observed by the Cluster spacecraft was about two times the ion inertia length \citep{chen08}. For B$_x$, we can assume to be a minor factor smaller than B$_{\rm 0}$ from our simulation result. Then, the minimum value of $|v_z|$ is a minor factor larger than $V_{\rm A}$. This is not a serious condition given the fact that electrons have a great chance of pre-acceleration at the X-line as we have already seen in Figure \ref{fig:HL-overview}a. Note also that, in the region away from the X-line but within the diffusion region, electrons can be bulk accelerated up to $V_{\rm Ae}$=$\sqrt{m_{\rm i}/m_{\rm e}}$V$_{\rm A}$ \citep[e.g.][]{hoshino01}.

\begin{figure}
\includegraphics[width=80mm]{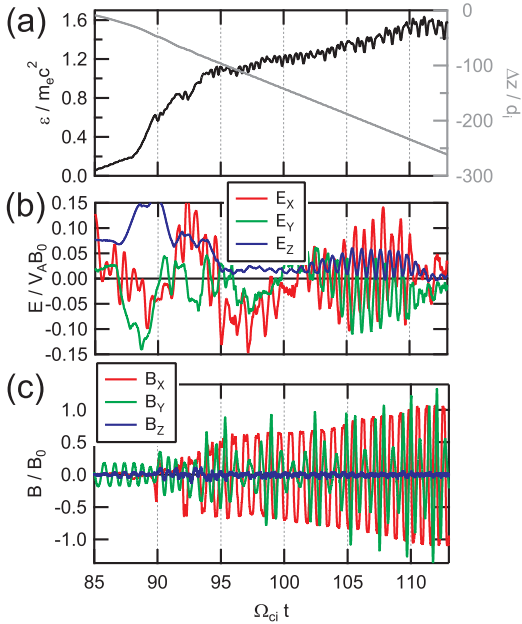}
\caption{Time profiles of (a) the energy and the displacement $\Delta$z, (b) the electric field and (c) the magnetic field felt by the particle during the trapping motion. In order to eliminate the particle noise, the electric field data have been smoothed by box average with the box size of $\Omega_{\rm ci}\Delta$t=2.7.\label{fig:HL-field}}
\end{figure}

An accelerated electron would be kept trapped as long as its gyro-radius r$_{\rm L}$  = $p_{\rm e}$/$eB_{l}$ 
% = m$_{\rm e}$v$_{e}$c/eB$_{\rm local}$ 
remains smaller than the size of the structure R where $p_{\rm e}$ is the electron momentum and B$_{l}$ is the local magnetic field magnitude felt by the particle. Since magnetic field lines of the inflowing plasma constantly accumulate on the island, we may assume that the size increases with the rate $\alpha_{\rm r}$V$_{\rm A}$ where $\alpha_{\rm r}$ is the instant reconnection rate. For the initial size of the magnetic island R$_{\rm 0}$ and the time from the start of the island expansion $\Delta$t, R = R$_{\rm 0}$ + $\alpha_r$V$_{\rm A}\Delta$t. Then, the trapping condition r$_{\rm L}<$ R can be rewritten as
\begin{equation}
%\frac{v_{\rm z}}{c} < 
\frac{p_{\rm e}}{m_{\rm e}c} <
\sqrt{\frac{m_{\rm i}}{m_{\rm e}}} \left(\frac{B_{l}}{B_0} \right) \left(\frac{\Omega_{\rm ce}}{\omega_{\rm pe}}\right) \left(\frac{R_{\rm 0}}{d_{\rm i}} + \alpha_r\Omega_{\rm ci}\Delta t \right) 
\label{eq:max}
\end{equation}
 but R$_{\rm 0}> $ d$_{\rm i}$, $\omega_{\rm pe}>\Omega_{\rm ci}$ and $B_l/B_0\sim$1 in our simulation and in the magnetotail as well so that the maximum velocity would be
\begin{equation}
v_{\rm z, max} \sim c
\end{equation}
in the non-relativistic regime, indicating `unlimited' electron acceleration by the `island surfing' mechanism (If the electron is actually energized up to a relativistic energy, equation (\ref{eq:max}) gives the upper limit). This efficient acceleration is an advantage of the closed field line geometry of the secondary islands.
%Once trapped in the magnetic island, the electron continues to gain energy and will never escape from the island unless the island is expelled from the diffusion region and/or the reconnection electric field decreases. Such `unlimited electron acceleration' occurs because the reconnection electric field accelerates electron in the $z$-direction and the electron current density increases. The increased current would then increase the magnetic field magnitude of the island, leading to larger Lorentz force. This is a positive feedback and, in a sense, the same as the electron self-reinforcing dynamics discussed recently \citep{wan08b}.

Note, however, that the `island surfing' works only for the island located within the diffusion region entirely covered by the reconnection electric field (Figure \ref{fig:HL-pots}c). Any other island outside the diffusion region lacks the reconnection electric field and has to generate motional electric field to accelerate electrons. This can only be achieved by a contracting motion (Figure \ref{fig:HL-pots}d). A bulk motion of the entire island does not accelerate electrons because it generates a positive out-of-plane electric field at one end and a negative out-of-plane electric field at the other end, resulting in a small, net energy gain. 

In our simulation, the electron in Figure \ref{fig:HL-overview} and \ref{fig:HL-field} has already reached $\varepsilon/m_{\rm e}c^2$=1 with the displacement of $\Delta z$/d$_{\rm i}\sim$150  during the initial interaction of the newly emerging island.  Note that we used an unrealistically high initial temperature $\varepsilon_{\rm th}$=9.2$\times$10$^{-3}$m$_{\rm e}c^2 \sim$ 4.7 keV due to limited computational resources. If we normalize simulated particle energies by $\varepsilon_{\rm th}$ which is $\sim$ 1 keV in the magnetotail and assume d$_{\rm i}$=500 km, the simulation indicates electron energization up to 110 keV with the dawnward displacement $\sim$12R$_{\rm E}$. Given the fact that the diameter of the magnetotail is $\sim$40R$_{\rm E}$,  the initial interaction with the emerging secondary island may explain  energetic ($<$127 keV) electrons observed by the Cluster spacecraft that passed through small scale magnetic islands \citep{chen08}.

%In our simulation, a fast motion ($\sim$0.3V$_{\rm A}$) of the island generated such electric field but, thanks to the reconnection electric field, the net energy gain remained positive. 

%Note that our measured value of $E_z\sim$0.025V$_{\rm A}$B$_{\rm 0}/c$ is somewhat smaller than the typical value, $E_z\sim$0.1V$_{\rm A}$B$_{\rm 0}/c$,  in a steady-state reconnection performed by PIC simulations. This is simply because our reconnection did not reach the steady-state.  
%Suppose a secondary island with a propagation speed of $\sim$0.3V$_{\rm A}$ was generated when $E_z\sim$0.1V$_{\rm A}$B$_{\rm 0}/c$ and the diffusion extended as long as 100d$_{\rm i}$, the time
%Ideally, the `island surfing' should be checked by a large scale PIC simulation perhaps with an aid of open boundary conditions.

The main part of the `island surfing' may work in much larger scale current sheets such as those found in the solar atmosphere. In this respect, we would also like to point out that magnetic reconnection in a spatially large current sheet (in the $y$-direction) would yield not just one magnetic island but a chain of islands \citep{loureiro07} so that a multitude of `island surfing' builds up to create a significant amount of energetic electrons.  In addition, if an external force existed to drive the magnetic reconnection, the reconnection electric field increases \citep{hoshino05} and the frequency of secondary island generation also increases \citep{wan08b}. Such a driven configuration is favorable for more energization by the `island surfing'.
%The `island surfing' becomes more important for the magnetic reconnection driven by an external force because such condition leads to the increase of the reconnection electric field \citep{hoshino05} as well as the number of secondary islands \citep{wan08b}. On the other hand, even in the Sweet-Parker type reconnection characterized by long current sheets, a chain of secondary islands can be formed \citep{loureiro07} so that a multitude of `island surfing' builds up to create a significant amount of energetic electrons. Finally,

We did not consider a guide field, but our preliminary simulation with a guide field (B$_g$=0.3B$_0$) also showed efficient electron energization within a secondary magnetic island, although its efficiency is still under investigation. %We anticipate that the `island surfing' mechanism may well be responsible for the energetic electrons observed in the Earth's magnetotail as well as the solar flares.

Finally, we would like to point out that the `island surfing' is not just a matter of energization of a few unusual electrons. It would give an insight toward the understanding of the roles of secondary islands in the context of  energy disspation during magnetic reconnection.

%%% End of body of article:

%%%%%%%%%%%%%%%%%%%%%%%%%%%%%%%%
%% Optional Appendix goes here
%
%%%%%%%%%%%%%%%%%
% Geophysical Research Letters only allows an appendix without a letter.
%% You can get this result with
%  \section*{Appendix}
%  or
%  \section*{Appendix: Title}
%%%%%%%%%%%%%%%%%
%
% \appendix resets counters and redefines section heads
% but doesn't print anything.
% After typing  \appendix
%
% \section{Here Is Appendix Title}
% will print
% Appendix A: Here Is Appendix Title
%
% \section*{Appendix}
% will print
% Appendix
%
% \section*{Appendix: Here Is Appendix Title}
% will print
% Appendix: Here Is Appendix Title
%
% For only 1 appendix \appendix \section{Appendix} is preferred.
% which will print
% Appendix A

%%%%%%%%%%%%%%%%%%%%%%%%%%%%%%%%%%%%%%%%%%%%%%%%%%%%%%%%%%%%%%%%
%
% Optional Glossary or Notation section, goes here
%
%%%%%%%%%%%%%%
% Glossary only allowed in Reviews of Geophysics
% \section*{Glossary}
% \paragraph{Term}
% Term Definition here
%
%%%%%%%%%%%%%%
% Notation -- End each entry with a period.
% \begin{notation}
% Term & definition.\\
% Second Term & second definition.
% \end{notation}
%%%%%%%%%%%%%%%%%%%%%%%%%%%%%%%%%%%%%%%%%%%%%%%%%%%%%%%%%%%%%%%%
%
%  ACKNOWLEDGMENTS

\begin{acknowledgments}
Simulations are performed on the SX-9 computer of ISAS, JAXA, Japan. We thank helpful comments from M. Swisdak.
\end{acknowledgments}

%% ------------------------------------------------------------------------ %%
%
%  REFERENCE LIST AND TEXT CITATIONS
%
% Either type in your references using
% \begin{thebibliography}{}
% \bibitem{}
% Text
% \end{thebibliography}
%
% Or,
%
% If you use BiBTeX for your References, please produce your .bbl
% file and copy the contents into your paper here.
%
% Follow these steps:
% 1. Run LaTeX on your LaTeX file.
%
% 2. Run BiBTeX on your LaTeX file.
%
% 3. Open the new .bbl file containing the reference list and
%   copy all the contents into your LaTeX file here.
%
% 4. Comment out the old \bibliographystyle and \bibliography commands.
%
% 5. Run LaTeX on your new file before submitting.
%
% AGU does not want a .bib or a .bbl file, but asks that you
% copy in the contents of your .bbl file here.

%Reference citation examples:

%...as shown by \textit{Kilby} [2008].
%...has been shown [\textit{Kilby et al.}, 2008].

%...as shown by \cite{jskilby}.
%...has been shown \citep{jskilbye}.

%% ------------------------------------------------------------------------ %%
%
%  END ARTICLE
%
%% ------------------------------------------------------------------------ %%

\end{article}

%% Enter Figures and Tables here:

% When submitting articles through the GEMS system:
% COMMENT OUT ANY COMMANDS THAT INCLUDE GRAPHICS.

% Figure captions go below this illustration; Table captions go above tables

% ONE-COLUMN figure/table, including eps graphics
%
% \begin{figure}
% \noindent\includegraphics[width=20pc]{samplefigure.eps}
% \caption{Caption text here}
% \end{figure}
% \end{document}
%
% \begin{table}
% \caption{}
% \end{table}
%
% ---------------
% TWO-COLUMN figure/table
%
% \begin{figure*}
% \noindent\includegraphics[width=39pc]{samplefigure.eps}
% \caption{Caption text here}
% \end{figure*}
%
% \begin{table*}
% \caption{Caption text here}
% \end{table*}
%
% see below for how to make landscape figures or tables

%%% End the article here:

\end{document}